\documentclass[aps,pra,superscriptaddress,amsmath,amssymb,twocolumn,floatfix,longbibliography]{revtex4-1}
\usepackage[pdftex]{graphicx}
\usepackage{bm}
\usepackage{bbold}
\usepackage[usenames,dvipsnames]{color}

\newcommand{\bg}{ \begin{gather} }
\newcommand{\eg}{\end{gather}}

\newcommand{\be}{ \begin{equation} }
\newcommand{\ee}{\end{equation}}

\newcommand{\corr}[1]{\langle #1\rangle}

\newcommand{\sgn}{\mathop{\rm sgn}}

\def\rprime{r'}

\begin{document}

\title{Ballistic correction to the density of states in an interacting three-dimensional metal}

\author{Daniil S. Antonenko}
\affiliation{Skolkovo Institute of Science and Technology, Skolkovo, Moscow Region, 143025, Russia}
\affiliation{L. D. Landau Institute for Theoretical Physics, Chernogolovka,
Moscow Region, 142432, Russia}
\affiliation{Moscow Institute of Physics and Technology, Moscow, 141700, Russia}

\author{Mikhail A. Skvortsov}
\affiliation{Skolkovo Institute of Science and Technology, Skolkovo, Moscow Region, 143025, Russia}
\affiliation{L. D. Landau Institute for Theoretical Physics, Chernogolovka,
Moscow Region, 142432, Russia}

\date{February 9, 2020}

\begin{abstract}
We study the tunneling density of states (DOS) in an interacting disordered three-dimensional metal and calculate its energy dependence in the quasiballistic regime, for the deviation from the Fermi energy, $E-E_F$, exceeding the elastic scattering rate.
In this region, the DOS correction originates from the interplay of the interaction and single-impurity scattering. Depending on the distance between the interaction point and the impurity, one should distinguish (i) the smallest scales of the order of the Fermi wavelength and (ii) larger spatial scales of the order of $\hbar v_F/|E-E_F|$, where $v_F$ is the Fermi velocity.
In two dimensions, the large-scale contribution prevails, resulting in a nearly universal DOS correction.
The peculiarity of Friedel oscillations in three dimensions is that the contributions from small and large scales are typically comparable, making the DOS correction sensitive to the details of the interaction and demonstrating a significant particle-hole asymmetry. On the other hand, we show that the non-analytic part of the DOS is determined by large scales and can be expressed in terms of the Fermi-surface characteristics only.
\end{abstract}

\maketitle

\section{Introduction}

Interplay of disorder and interaction in electron systems continues to be a hot topic in condensed matter physics for decades.
Among the milestones are the profound independence of the superconducting critical temperature on disorder (Anderson's theorem \cite{Anderson_ATh, AbrikosovGorkov_ATh1, AbrikosovGorkov_ATh2}), theory of superconducting fluctuations \cite{VarlamovLarkin_book}, interaction correction to the conductivity \cite{AltshulerAronovLee}, theory of dephasing in dirty metals \cite{AltshulerAronovKhmelnitsky_dephasing}, Fermi liquid theory for diffusive metals \cite{Fin83,Fin,BelitzKirkpatrick94}, etc.

Influence of electron-electron interaction on the one-particle density of states (DOS) in disordered metals was studied in the pioneering work by Altshuler and Aronov \cite{AltshulerAronov_first, AltshulerAronov_DOS_ZBA_1979}. They derived the perturbative in the interaction strength correction to the tunneling DOS, assuming electron motion to be diffusive.
The resulting expression for the DOS anomaly is singular at the Fermi energy, $E_F$, leading to the famous logarithmic behavior in two dimensions (2D). In the three-dimensional (3D) case, the singularity is weaker:
\be
\label{conventional_diffusive}
\frac{\delta \nu_\text{diff}(E)}{\nu_0} 
= 
\frac{c_\text{d} \lambda \sqrt{|E-E_F| \tau }}{\left(k_F l \right)^2}  \, 
, \quad \:\: |E-E_F| \ll \frac{1}{\tau} ,
\ee
where $\nu_0$, $\tau$, and $l = v_F \tau$ are the noninteracting DOS (per one spin projection), elastic scattering time, and elastic mean free path at $E_F$, $\lambda$ is the dimensionless interaction strength, and $c_\text{d}$ is a numeric constant, depending on the spatial profile of the interaction ($\hbar=1$).

Theoretical analysis of Altshuler and Aronov was followed by a number experimental studies that confirmed their results \cite{AltshulerAronov_book_1985} in the 2D \cite{WhiteGarno_1985_CorrectionTo2DimDos} and 3D geometries \cite{McMillanMochel, HertelBishopDynes_1983_NbSi}, and in the crossover between them \cite{ImryOvadyahu_1982}. 
The first-order correction was later extended to a non-perturbative level in Refs.\ \cite{Fin,LS,KA}.
With the increase of disorder, the Altshuler-Aronov correction gets transformed into a fully developed Coulomb gap at the insulating side of the metal-insulator transition \cite{LeeShklovskii_CoulombGapSemicondTunneling_1999, BokachevaDynes_2004_GdSi}. 

The interaction correction was initially expected to be cut off at the border of the \emph{diffusive region}, at
$|E-E_F| \sim 1/\tau$. 
However, as has been shown later by Rudin, Aleiner, and Glazman \cite{RudinAleinerGlazman_TunnelingZBQuasiballistic_1997}, 
this correction actually extends to the \emph{quasiballistic region}, $|E-E_F| \gg 1/\tau$, where it originates from scattering on an effective potential produced by Friedel oscillations around a single impurity. In the 2D geometry, this effect leads to the logarithmic dependence of $\delta\nu(E)$,  similar to the behavior in the diffusive region.
An attempt to generalize this finding to the 3D geometry was made by Koulakov \cite{Koulakov_QuasiballisticDOS_2000}, who considered Fock (exchange) contribution in a number of simplifying approximations and predicted a linear behavior of the correction in the quasiballistic regime:
\color{black}
\be
\label{Koulakov-result}
\frac{\delta \nu_\text{ball}^\text{(Koul)} (E)}{\nu_0} \sim \frac{\lambda}{k_F l} \frac{|E-E_F|}{E_F}, \quad \:\: \frac{1}{\tau} \ll |E-E_F|.
\ee

\begin{figure}
\vspace{-3mm}
	\includegraphics[width=1.0\linewidth]{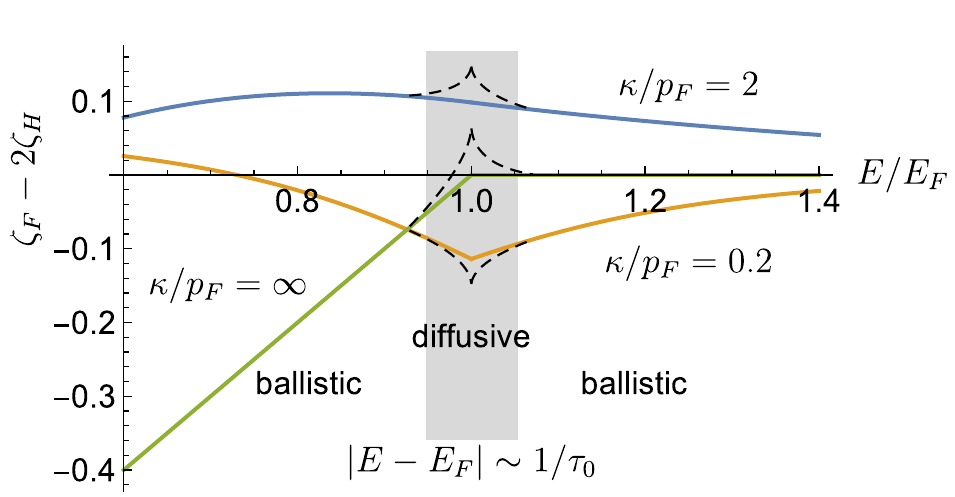}
	\protect\caption{
Energy dependence of the quasiballistic correction~\eqref{delta_nu_zeta} to the tunneling DOS for the Yukawa interaction \eqref{Coulomb} at different screening wave vectors $\kappa/p_F=0.2$, 2, $\infty$ (point-like interaction). Dashed lines are a sketch of the diffusive square-root contribution \eqref{conventional_diffusive} in the vicinity of the Fermi energy.} 
\label{fig:different_kappa}
\end{figure}

In the present paper we reconsider the quasiballistic contribution to the DOS in a 3D metal with a short-range interaction. We take into account the Hartree diagram neglected in Ref.\ \cite{Koulakov_QuasiballisticDOS_2000} and accurately trace the energy dependence of the correction in the whole ballistic region $|E-E_F| \gg 1/\tau$. As expected for ballistic systems, the behavior of the DOS correction is sensitive to the details of both interaction and dispersion relation. The obtained dependence of $\delta\nu(E)$ is typically asymmetric with respect to the Fermi energy. This asymmetry is most pronounced in the case of point-like interaction and parabolic spectrum, when the DOS correction completely vanishes for $E>E_F$. In general, the correction can be conveniently represented in the form
\be
\label{delta_nu_zeta}
	\frac{\delta\nu_\text{ball}(E)}{\nu_{0}} = 
	\frac{\pi\lambda}{4p_{E}l} 
  [\zeta_F(E) - 2 \zeta_H(E)],
\ee 
where $p_E$ is the momentum corresponding to the energy $E$ for a given dispersion relation, and $\zeta_F(E)$ and $\zeta_H(E)$ are the dimensionless contributions of the Fock (exchange) and Hartree diagrams, both being of the order of 1. In the case of parabolic dispersion and Yukawa interaction potential, these functions are shown in Fig.~\ref{fig:different_kappa} for different screening radii.

While the behavior of $\delta\nu_\text{ball}(E)$ essentially depends on the dispersion relation in the whole band, we obtain that the jump of its derivative at $E_F$ (clearly seen in Fig.~\ref{fig:different_kappa}) is universal in a sense that it is determined solely by the system properties at the Fermi surface and interaction potential in the momentum representation:
\be
\label{jump}
  \left. 
    \frac{\partial [ \delta\nu_\text{ball}(E)/\nu_0]}{\partial(p_E/p_F)} 
  \right|_{E_F - 0}^{E_F + 0}  
  = \frac{\pi\nu_0[V(0) - 2 V(2p_F)]}{k_F l}  ,
\ee
where the two terms correspond to the Fock (exchange) and Hartree contributions, respectively.
It is this jump that was actually obtained by Koulakov \cite{Koulakov_QuasiballisticDOS_2000}, who took into account only the contribution from the Fermi surface and therefore was able to get only the non-analytic part of the ballistic correction.
Indeed, the jump calculated from the Koulakov's result \eqref{Koulakov-result} exactly coincides with Eq.~\eqref{jump}, provided that the Hartree contribution is neglected.

In terms of spatial scales, the DOS correction (\ref{delta_nu_zeta}) contains contributions both from small scales $r\sim 1/k_F$ and large scales $r\sim v_F/|E-E_F|$, where $r$ is the distances between the interaction point and the impurity. Therefore it is not universal. 
However, the derivative jump comes from large spatial scales $r \gg 1/k_F$, leading to a compact formula (\ref{jump}).

The paper is organized as follows. In Sec.~\ref{sec:DOS_evaluation} we present the general expression for the lowest diagrammatic contribution to the DOS in the quasiballistic limit. 
The simplest case of the point-like interaction is analyzed in Sec.~\ref{S:point-like}, both for parabolic and non-parabolic dispersion. The influence of a finite interaction range is addressed in Sec.\ \ref{S:Yukawa} for a model case of the Yukawa interaction.
 Universality of the derivative jump is proven in Sec.\ \ref{S:Jump}. Results are summarized in Sec.\ \ref{S:conclusion}. 
The Appendix highlights the difference between 2D and 3D Friedel oscillations.

\section{Interaction correction in the quasiballistic regime}
\label{sec:DOS_evaluation}


We consider a spin-1/2 electron gas interacting with a weak pairwise potential $V(\bm{r})$
subject to an impurity scattering described by the random potential $U(\bm{r})$. The latter is assumed to be short-range and is described by the standard model of a Gaussian white noise specified by the correlation function $\corr{U(\bm{r})U(\bm{r}')}=\delta(\bm{r}-\bm{r}')/(2\pi\nu_0\tau)$. 
The analysis will be performed in the zero-temperature limit, $T=0$.

The tunneling DOS can be expressed in terms of the casual (Feynman) single-particle Green function taken at coinciding points: $\nu(E)=-\sgn(E-E_F)\mathop{\rm Im}G_E(\bm{r}_0,\bm{r}_0)/ \pi$.
The latter can be written in terms of the retarded and advanced Green functions as
$G_{E} = G_{E}^{R(A)}$ for $E>E_F$ ($E<E_F$).

\begin{figure}
	\includegraphics[width=0.8\linewidth]{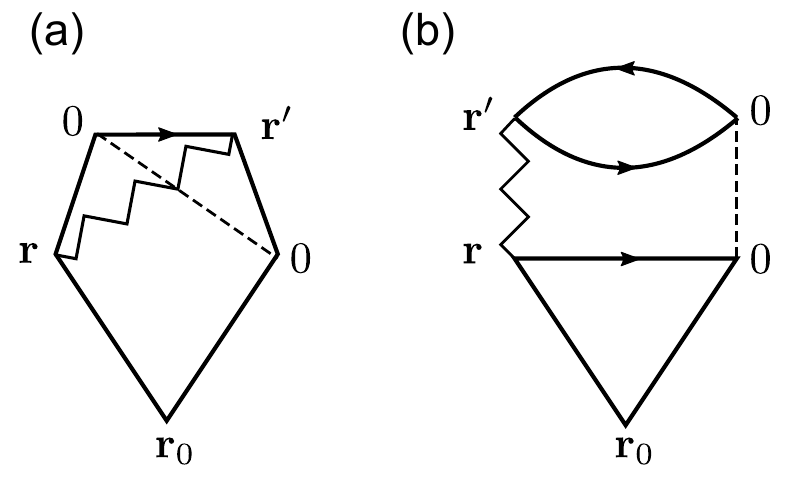}
	\protect\caption{(a) Fock (exchange) and (b) Hartree diagrams for the average DOS (measured at point $\bm{r}_0$) of a disordered interacting metal in the ballistic energy window $|E-E_F|\gg1/\tau$. Here solid, zigzag, and dashed lines represent electron Green function, electron-electron interaction, and disorder potential correlation function, respectively. } 
\label{fig:Diagramms}
\end{figure}

In the quasiballistic region, $|E-E_F| \gg 1/\tau$, correction to the tunneling DOS is given by the diagrams shown in Fig.~\ref{fig:Diagramms} (both have symmetric counterparts) \cite{RudinAleinerGlazman_TunnelingZBQuasiballistic_1997}. These are the processes of the lowest order in the interaction and disorder potential [in the quasiballistic energy range, the diffusive ladder would transfer a typical momentum $q \gtrsim \min (p_F, \kappa) \gg 1/l$ and therefore it does not appear] .

Evaluation of the diagrams is more convenient in the real-space representation.
The resulting correction to the average DOS can be written as
\be
\label{delta_nu_original_integral}
	\frac{\delta \nu (E)}{\nu_0} = \frac{\sgn(E-E_F)}{\pi^{2}\nu_0^{2}\tau} \mathop{\rm{Re}} 
	\int d \bm{r} \, d \bm{r}' \, V_{\left| \bm{r} - \bm{r} ' \right|} 
	 ( \eta_F - 2 \eta_H ),
\ee
where the Fock (exchange) and Hartree contributions read
\begin{gather}
\label{eta_F}
	\eta_F(E) = \int\frac{dE'}{2\pi}  \left[G_{E}G_{E}\right](r) \, G_{E}(\rprime) G_{E'}(r) G_{E'}(\rprime), \\
\label{eta_H}
	\eta_H(E) = \int\frac{dE'}{2\pi} \left[G_{E}G_{E}\right](r) \, G_{E}(r) G_{E'}(\rprime) ^2 . 
\end{gather}
Here the object $[G_{E}G_{E}]$ is a convolution of two Green functions over the expectation point $\boldsymbol{r}_0$:
\be
\left[G_{E} G_{E} \right] = \int G_E(\bm{\rho}) G_E(\bm{r}-\bm{\rho}) \, d \bm{\rho} 
  = 
- \frac{\partial G_E(\bm{r})}{\partial E} .
\ee

As usual \cite{AGD}, integration over $E'$ in Eqs.~\eqref{eta_F} and \eqref{eta_H} should be regularized by introducing the factor $e^{+i0 E'}$, which allows one to close the integration contour in the upper half-plane. Using the analytic properties of $G^R$, we can replace $\eta_F$ and $\eta_H$ by equivalent expressions 
\begin{multline}
	\label{analytical_continuation_F}
	{\eta}_F = \int^{E_F}_{0} \frac{dE'}{2\pi}  \left[G_{E}G_{E}\right](r) \, G_{E}(r') 
\\ 
	{} \times \big[ G_{E'}^A(r) G_{E'}^A(r') - G_{E'}^R(r) G_{E'}^R(r') \big], 
\end{multline}
\vspace{-10pt}
\begin{equation}
	\label{analytical_continuation_H}
	{\eta}_H = \int^{E_F}_{0} \frac{dE'}{2\pi}  \left[G_{E}G_{E}\right](r) \, G_{E}(r) \big[ G_{E'}^A(r')^2  - G_{E'}^R(r')^2 \big] ,
\end{equation}
Initially energy integration in Eqs.~\eqref{analytical_continuation_F} and \eqref{analytical_continuation_H} is performed from $-\infty$ to $E_F$,
however since at negative energies $G^R=G^A$ (vanishing DOS), the lower limit was replaced by 0.

The general expression for the quasiballistic correction determined by Eqs.~\eqref{delta_nu_original_integral}, \eqref{analytical_continuation_F} and \eqref{analytical_continuation_H} involves integrals over $\bm{r}$, $\bm{r}'$, and $E'$, which cannot be evaluated in a closed form for an arbitrary spectrum and interaction potential. To get insight on the behavior of $\delta\nu_\text{ball}(E)$, we perform calculations in the case of the point-like interaction (Sec.~\ref{S:point-like}) and screened Yukawa potential (Sec.~\ref{S:Yukawa}), where the result can be obtained in a closed analytic form.

\section{Point-like interaction}
\label{S:point-like}

In this section we address the simplest case of a point-like interaction $V(\bm{r}) = \lambda \delta(\bm{r})/ 2 \nu_0$. Though it formally corresponds to the $\kappa \rightarrow \infty$ limit of the Yukawa potential \eqref{Coulomb} considered in Sec.\ \ref{S:Yukawa}, it is instructive to consider it separately in order to demonstrate 
the role of different scales and compare with the results of Refs.\  \cite{RudinAleinerGlazman_TunnelingZBQuasiballistic_1997} and \cite{Koulakov_QuasiballisticDOS_2000}.

\color{black}

\subsection{Parabolic dispersion}

For the parabolic dispersion, $E(p)=p^2/2m$, the unperturbed 3D causal Green function in the coordinate representation has the form:
\be
\label{Green_function}
	G_E(r)=-\frac{\pi \nu_0}{p_F r}e^{ i r p_E \cdot \text{sgn} \left( E - E_F \right)},
\ee
where $p_E = \sqrt{2 m E}$. 

Substituting Eq.\ \eqref{Green_function} into Eqs.\ 
\eqref{delta_nu_original_integral}, \eqref{analytical_continuation_F} and \eqref{analytical_continuation_H}, one immediately obtains the correction in the form \eqref{delta_nu_zeta} with equal contributions:
\be
\label{delta_nu_sign_diff}
	\zeta_F(E) = \zeta_H(E) = 
	\frac{1}{\pi}  \int_{0}^{E_{F}}\frac{dE'}{E_{F}} 
\int_{0}^{\infty} dr \, \frac{\sigma_{EE'}(r)}{r} ,
\ee
where we have introduced 
\be
\label{sigma-def}
  \sigma_{EE'}(r) = 
  \sin 2 \left( p_E - p_{E'} \right) r 
- \sin 2 \left( p_E + p_{E'} \right) r 
.
\ee
Integration over $r$ in Eq.\ \eqref{delta_nu_sign_diff} is expressed in terms of the Dirichlet integral, leading to $\zeta_F = \zeta_H = \zeta$, where
\be
\label{point-like_Ep_integral}
\zeta(E) =
\frac{1}{2}  \int_{0}^{E_{F}}\frac{dE'}{E_{F}}
\left[\text{sgn}\left(E - E' \right)  -  \text{sgn}\left( E + E' \right)\right] .
\ee
Performing now trivial integration over $E'$, one gets
\be
\label{point_like_result}
\zeta_F(E) - 2 \zeta_H(E) = \frac{E - E_F}{E_F} \theta_{E_F - E},
\ee
Note that, somewhat counterintuitively, strongly oscillating, $\int dr \, (1/r) \sin 2 (p_E + p_{E'})r$, and slowly oscillating, $\int dr \, (1/r)\sin 2 (p_E - p_{E'}) r$, integrals [see Eqs.\ \eqref{delta_nu_sign_diff} and \eqref{sigma-def}] give exactly the same contributions and cancel each other for energies $E > E_F$, so that the correction completely vanishes above the Fermi energy (see Fig.~\ref{fig:different_kappa}).

The jump of the derivative of $\delta\nu_\text{ball}(E)$ at the Fermi energy is given by Eq.\ \eqref{jump} with $V(q) = \lambda/2\nu_0$.

\subsection{Non-parabolic dispersion}
\label{sec:nonparabolic_dispersion}

A remarkable feature of Eq.\ \eqref{point_like_result} obtained for the parabolic spectrum is the absence of the correction at $E > E_F$. To study the robustness of this result, we examine now the effect of spectral unharmonicity on the form of the DOS correction, considering parabolic spectrum perturbed by a quartic nonlinearity:
\be
\label{unharmonicity_spectrum}
	E(p) = ( p^2 + \alpha^2 p^4 )/2m.
\ee
In this case the Green function can also be written explicitely:
\be
\label{Green_function_modified}
	G_E(r)=-\frac{\pi \nu_E}{p_E r} \left[  e^{ i r p_E \cdot \text{sgn} \left( E - E_F \right)}  -  e^{- r \tilde{p}_E }   \right],
\ee
where $\nu_E$ is the density of states at energy $E$, $p_E$ now is the positive root of the equation $E(p) = E$, and $\tilde{p}_E = \sqrt{p_E^2 + 1/\alpha^2}$.

\begin{figure}
	\includegraphics[width=1.0\columnwidth]{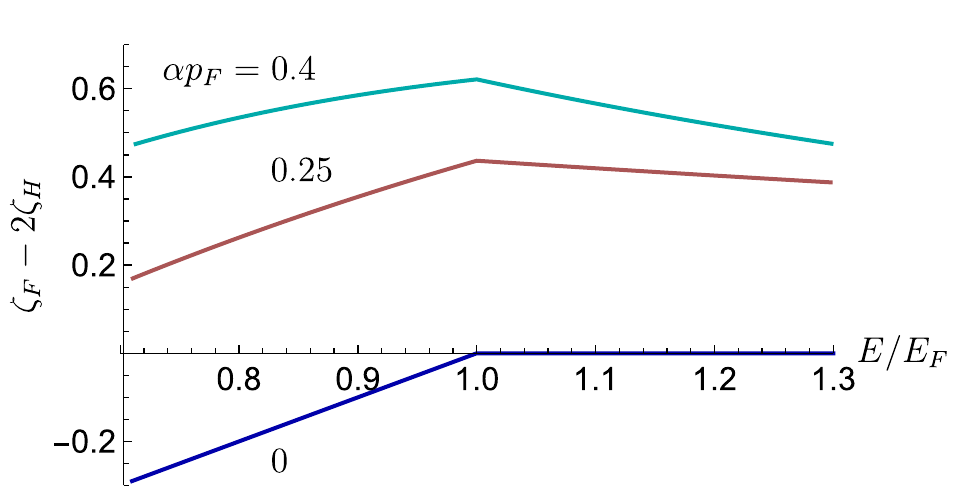}
	\caption{Energy dependence of the quasiballistic DOS correction \eqref{delta_nu_zeta} for the point-like interaction ($\kappa/p_F=\infty$)
in the presence of the quartic unharmonicity of the spectrum [Eq.~(\ref{unharmonicity_spectrum})] with $\alpha p_F = 0$ (parabolic dispersion), 0.25, 0.4.}
	\label{fig:nonlinear_perturb}
\end{figure}

Substituting Eq.\ \eqref{Green_function_modified} into Eqs.\ \eqref{analytical_continuation_F} and \eqref{analytical_continuation_H}, we obtain the DOS correction as a function of energy and parameter $\alpha$. Since the analytical answer is too lengthy, we present the results graphically in Fig.~\ref{fig:nonlinear_perturb}, where the lower curve corresponds to the parabolic dispersion ($\alpha=0$). We see that the main effect of unharmonicity is to make $\delta\nu_\text{ball}(E)$ finite at $E>E_F$, whereas the presence of the cusp at $E=E_F$ is robust with respect to spectrum deformation. Calculations (omitted here) show that the derivative jump is still given by Eq.\ \eqref{jump}, thus illustrating its universality when expressed in terms of the Fermi surface properties. This statement will be proven in Sec.~\ref{S:Jump}.%

\subsection{Relevant spatial scales in 2D and 3D}
\label{sec:point-like_discussion}

In this section we focus on the difference between Friedel oscillations in the 2D and 3D geometries, discuss the relevant spatial scales, and explain the origin of nonuniversality of the 3D ballistic correction contrary to the universal correction in two dimensions  \cite{RudinAleinerGlazman_TunnelingZBQuasiballistic_1997}.

In three dimensions, the integrand in Eq.\ \eqref{point-like_Ep_integral} is linear in $E'$, with the contribution of both terms being of the same order (for parabolic dispersion the total integrand is nonzero only in the strip $E < E' < E_F$ due to the cancellation of the two terms, but for an arbitrary dispersion there is no such cancellation). 
Since the first (second) term in Eq.\ \eqref{point-like_Ep_integral} corresponds to large (small) scales, that means that in the 3D case both large scales (Fermi surface properties) and small scales $r\sim 1/k_F$ (nonuniversal behavior, energies deep in the Fermi sea) are equally important. Such a non-universality is illustrated by the results of Sec.\ \ref{sec:nonparabolic_dispersion} presented in Fig.~\ref{fig:nonlinear_perturb}. 

In two dimensions, slower decay of Friedel oscillations (for the details, see Appendix \ref{app:2D}) leads to a more singular behavior of the energy integrand in the expression for $\zeta=\zeta_F=\zeta_H$ compared to that given by Eq.\ \eqref{point-like_Ep_integral}:
\be
\label{delta_nu_2D_singular_main_part}
	\zeta^\text{(2D)} =
	\frac{p_F}{\pi^2}
	\int_{0}^{E_F}\frac{dE'}{E_F} 
	\left( \frac{1}{p_E - p_{E'}} + \frac{1}{p_E + p_{E'}}  \right),
\ee
In the leading approximation, the second term originating from small scales $r\sim 1/k_F$ can be neglected, and the energy integral becomes logarithmic. In terms of spatial scales, it corresponds to a broad range of distances $1 \ll k_Fr \ll E_F/|E-E_F|$, where the Green functions can be replaced by their asymptotical expressions, determined by the Fermi surface properties. The same reasoning justifies the stationary phase analysis of scattering on the Friedel oscillations used in Ref.~\cite{RudinAleinerGlazman_TunnelingZBQuasiballistic_1997}. Altogether this makes a quasiballistic correction to the density of states be determined by the Fermi surface and symmetric with respect to $E_F$.

\section{Yukawa potential and parabolic dispersion}
\label{S:Yukawa}

To illustrate the role of the interaction range, in this section we consider the model with the electron-electron interaction described by the Yukawa potential:
\be
\label{Coulomb}
	V_{\bm{r}} = \frac{\lambda}{4\pi} \frac{\kappa^2}{2 \nu_0} \frac{ e^{- \kappa r}}{r} ,
\ee
where $\lambda\ll1$ is a dimensionless interaction constant and $\kappa$ is an arbitrary screening wave vector. 

In this model, the coordinate integrals in Eqs.\
\eqref{analytical_continuation_F} and \eqref{analytical_continuation_H} can be analytically calculated in the variables $|\bm{r}|$, $|\bm{r}'|$, and $|\bm{r} - \bm{r}'|$. Firstly, we take integrals over  $|\bm{r} - \bm{r}'|$ and $|\bm{r}|$ arriving for the following expressions for the Fock contribution in Eq.\ \eqref{delta_nu_zeta}:
\be
\label{zeta_F}
	\zeta_F(E) =  \int_{0}^{E_F}  \frac{dE'}{E_F}  \left[ F(p_E + p_{E'}) - F(p_E - p_{E'}) \right] , 
\ee
where
\be
		F(p) = \frac{1}{\pi} \frac{\kappa^2}{p^2 + \kappa^2} \mathop{\rm arccot} \frac{p}{\kappa}.
\ee
The Hartree contribution is given by 
\be
	\label{zeta_H}
	\zeta_H(E) = \frac{ \kappa^2}{(2 p_E)^2 + \kappa^2} \left[ \left(1 - \frac{E}{E_F} \right) \theta_{E_F - E}  + \beta_{2p_F /\kappa} 
	\right] ,
\ee
with
\be
  \beta_x = \frac{2}{\pi x} \left[ 1 - \frac{1 + x^2}{x} \arctan x \right] . 
\ee
Though the energy integral in Eq.\ \eqref{zeta_F} can be done analytically, we left it  unevaluated for compactness.
For weak screening ($\kappa \ll p_F$), the Hartree contribution is much smaller than the Fock one (as in the diffusive limit).

\begin{figure}
	\includegraphics[width=1.0\linewidth]{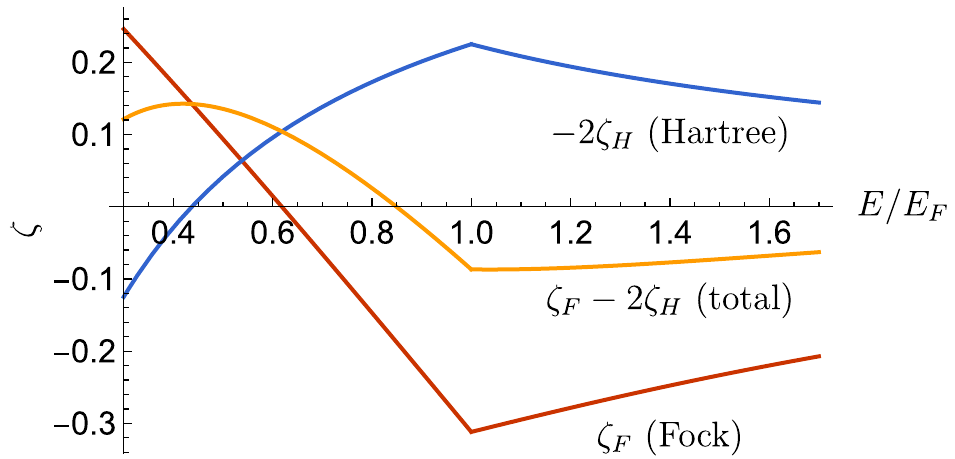}
	\protect\caption{Energy dependence of the Fock, Hartree, and total contributions to 
$\delta\nu_\text{ball}(E)$ [see Eq.\ \eqref{delta_nu_zeta}]
for the Yukawa interaction with the screening parameter $\kappa = p_F$.
}
\label{fig:Coulomb_Hartree_and_Fock}
\end{figure}

The total correction $\zeta_F(E)-2\zeta_H(E)$ for several values of $\kappa/p_F$ is shown in Fig.\ \ref{fig:different_kappa}.
At $\kappa / p_F = 1$, the results for $\zeta_F(E)$ and $\zeta_H(E)$ together with the total contribution are presented in Fig.~\ref{fig:Coulomb_Hartree_and_Fock}.

Both the Fock and Hartree contributions exhibits a jump of the derivative at $E_F$ as given by Eq.\ \eqref{jump} with $V(q) = (\lambda/2\nu_0) \kappa^2/(q^2+\kappa^2)$.
The sign of this jump is sensitive to the relation between $\kappa$ and $p_F$: The ballistic DOS correction exhibits a dip at $\kappa < 2 p_F$, which turns to a cusp at $\kappa > 2 p_F$, see Fig.\ \ref{fig:different_kappa}.

Our results for the DOS correction refer to the ballistic energy range, $|E-E_F|\gg1/\tau$ [in that sense the infinitesimal zero in Eq.\ \eqref{jump} should be understood]. In the diffusive limit, $|E-E_F|\ll1/\tau$, the correction is described by the Altshuler-Aronov expression \eqref{conventional_diffusive}, with the sign being determined by that of $V(0) - 2 \left\langle V(\boldsymbol{p} - \boldsymbol{p}^\prime) \right\rangle$, where the angle brackets in the Hartree term denote averaging over $\boldsymbol{p}$ and $\boldsymbol{p}^\prime$ lying on the Fermi surface \cite{AltshulerAronov_book_1985}. In the model of Yukawa potential, $\delta\nu_\text{diff}(E) \propto 1-2(\kappa/2p_F)^2\log[1+(2p_F/\kappa)^2]$, hence the diffusive DOS correction shows a cusp at $\kappa < 1.26\, p_F$ and a dip at $\kappa > 1.26\, p_F$. The behavior of the diffusive contribution is sketched by dashed lines in Fig.\ \ref{fig:different_kappa}.

\section{Universality of the derivative jump at the Fermi energy}
\label{S:Jump}

The quasiballistic correction to the DOS in the 3D geometry is non-universal, depending both on the details of the interaction potential $V(r)$ and on the electron dispersion in the whole band, which is \emph{a priori}\/ unknown. 
However it appears that the jump of $\partial\nu_\text{ball}(E)/\partial E$ at the Fermi energy is determined only by the Fermi surface properties and interaction-dependent constants, as given by Eq.\ \eqref{jump}.
In Secs.\ \ref{S:point-like} and \ref{S:Yukawa} we have already observed this fact in those cases when $\nu_\text{ball}(E)$ can be evaluated analytically.

Below we prove the universality of the jump for an arbitrary dispersion relation and short-range potential. First, we show how that can be verified for the point-like electron-electron interaction without relying on the exact analytic solution, and then demonstrate it in the general case.

For point-like interaction with the parabolic dispersion, $\partial\nu_\text{ball}(E)/\partial E$ can be obtained by differentiating  $\sigma_{EE'}$ in Eq.\ \eqref{delta_nu_sign_diff}. That removes the factor $r$ in the denominator, and the integral over $r$ yields $\delta(p_E - p_{E'})$ [an analogous term $\delta(p_E + p_{E'})$ does not contribute]. 
The $\delta$-peak is to be integrated over the region $0 < E' < E_F$. Therefore it contributes to the integral at $E<E_F$ and does not contribute at $E>E_F$. This results in the jump of the derivative given by Eq.\ \eqref{jump}. 
An important observation is that this $\delta$-peak originates from large distances, where the Green functions can be replaced by their asymptotics. The latter is determined by the vicinity of the Fermi surface and, consequently, so is the derivative jump.

This reasoning can be easily generalized to an arbitrary finite-range static electron-electron interaction $V_r$. As we have seen above, the jump of $\partial\nu_\text{ball}(E)/\partial E$ is associated with $\delta(p_E - p_{E'})$, which originates from large distances between the impurity location 0 and interaction points $r$ and $r'$. This circumstance allows us to use the asymptotic expression for the Green function,
$G_E^{R(A)}(r) \sim - [\pi \nu(E) ] / (p_E r) e^{\pm i r p_E}$, expressed solely in terms of the Fermi surface properties.
Substituting it into Eqs.~\eqref{delta_nu_original_integral}, \eqref{analytical_continuation_F}, and \eqref{analytical_continuation_H}, and extracting the jump as explained above, we arrive at
\be
\label{jump-2}
  \left. 
    \frac{\partial [ \delta\nu_\text{ball}(E)/\nu_0]}{\partial(p_E/p_F)} 
  \right|_{E_F - 0}^{E_F + 0}  
  = \frac{\pi(\lambda_F - 2 \lambda_H)}{2k_F l}  ,
\ee
with the coefficients $\lambda_F$ and $\lambda_H$ given by
\begin{gather}
\label{lf}
  \lambda_F 
  = 
  \frac{\nu_0}{2\pi^2} 
  \hat{\mathfrak D} \int d\bm{r} \, d\bm{r}' \,
  V_{\bm{r} - \bm{r}'} \frac{r + \rprime}{r r'^2} 
  e^{ i(p_E - p_{E'})(r+r')},
\\
\label{lh}
  \lambda_H 
  = 
  \frac{\nu_0}{\pi^2} 
  \hat{\mathfrak D} \int d\bm{r} \, d\bm{r}' \, 
  V_{\bm{r} - \bm{r}'} \frac{1}{r'^2} 
  e^{ 2i p_E r - 2 i p_{E'}r'}.
\end{gather}
Here the operator $\hat{\mathfrak D}$ extracts the coefficient in front of $\delta(p_E-p_{E'})$, sets $E=E'=E_F$, and takes the real part of the resulting expression.
To calculate the integrals \eqref{lf} and \eqref{lh}, we switch to new variables $\bm{r}$ and $\bm{\rho} = \bm{r}' - \bm{r}$.
As relevant $\rho \ll r$ at $r\rightarrow \infty$ due to the finite interaction range, we put $\rprime \rightarrow r$ in the prefactor and expand $\rprime = r + (\bm{r}\cdot\bm{\rho})/r$ in the exponent. Then, taking the integrals over $\bm{r}$ and $E_F$ we arrive at Eq.\ \eqref{jump} for the derivative jump.

\section{Discussion and conclusion}
\label{S:conclusion}

To sum up, we have studied the tunneling DOS correction caused by the collective impact of electron-electron interaction and short-range disorder in the quasiballistic energy range $|E-E_F| > 1/\tau$. In the 3D case, this correction has two comparable contributions, one coming from the Fermi wave-length scale $k_F^{-1}$ and another coming much larger scales $\sim v_F/|E-E_F|$.
As a result, $\delta\nu_\text{ball}(E)$ becomes  model-dependent and asymmetric with respect to the Fermi energy. At the same time, the jump of the derivative of $\delta\nu_\text{ball}(E)$ at $E_F$ is determined by large scales and therefore is less sensitive to the microscopic details. According to Eq.\ \eqref{jump}, it is expressed in terms of the Fermi surface characteristics and Fourier components of interaction potential at $q=0$ and $q=2p_F$. That somehow resembles the situation in 2D, where, however, the whole logarithmic DOS correction in the quasiballistic region is proportional to $V(0)-2V(2p_F)$.

The jump of the derivative of $\delta\nu_\text{ball}(E)$ at $E_F$ in the 3D case was first obtained by Koulakov \cite{Koulakov_QuasiballisticDOS_2000}. He  considered only the Fock diagram and neglected all terms responsible for smooth analytic energy behavior near $E_F$. Our analysis demonstrates that his account of large spatial scales was correct. At the same time, in the 3D geometry short scales of the order of the Fermi wave length give a comparable analytic contribution, which is strongly model-dependent. The interplay between the short-scale and large-scale contributions results in a significant asymmetry of $\delta\nu_\text{ball}(E)$ that is most pronounced for the point-like interaction and parabolic dispersion, when the DOS above the Fermi energy remains undisturbed.

Interestingly, a similar $|E - E_F|$ nonanalyticity of the DOS is known for 2D interacting systems, where it stems from the plasmon mode (pole at $q \sim \omega^2 / \kappa v_F^2$) \cite{Khveshchenko} and electron-hole excitations (singularity at $q \sim |\omega| / v_F$) \cite{Mishchenko_Andreev}, where $\omega=E-E'$. As it does not contain a small $1/k_Fl$ factor, this correction may even exceed the result \eqref{2D_logarithm} of Rudin \emph{et~al.}~\cite{RudinAleinerGlazman_TunnelingZBQuasiballistic_1997}. However both mechanisms studied in Refs.\ \cite{Khveshchenko, Mishchenko_Andreev} are ineffective in the 3D geometry, since the corresponding processes with low energy transfer $|\omega| \ll E_F$ are either absent (3D plasmons are gapped) or suppressed due to the reduced phase space in 3D (electron-holes pairs). 

In real 3D materials, electrons interact by the dynamically screened Coulomb interaction rather than by the static short-range potential. However, in studying the ballistic contribution to the \textit{nonanalytical} part of the DOS correction caused by electron scattering on 3D Friedel oscillations, one can safely use the static approximation, as the relevant energy transfer is small,
while the momentum transfer is large. 
This statement was checked by Koulakov \cite{Koulakov_QuasiballisticDOS_2000}, who observed that the derivative jump for the ballistic part of the DOS correction is the same both for the static short-range potential and the dynamically screened Coulomb interaction in the random phase approximation.
Note however that our model cannot formally describe the Coulomb interaction, since the latter corresponds to $\lambda\sim1$, where the perturbation theory is not valid. Instead one can try a usual expansion in a small gas parameter $r_s = e^2/v_F \sim (\kappa / p_F)^2$ \cite{Mahan, Kozii}, which makes the Hartree contribution negligible compared to the Fock one both in the diffusive and ballistic regimes (see discussion in the end of Sec.~\ref{S:Yukawa}).

Direct experimental observation of the ballistic correction as given by Eq.~\eqref{delta_nu_zeta} is complicated since its contribution to the change of the slope $\partial \nu(E) / \partial E$ is parametrically smaller than that of the unperturbed 3D DOS $\nu_0(E)$ by the factor $1/(k_F l)$. 
However the latter is an analytical function of energy in the vicinity of $E_F$, which makes the universal derivative jump of the interaction correction \eqref{jump} accessible for experimental determination. 
Finally, the obtained DOS correction can be important in calculation of physical quantities that depend on the DOS on a large energy scale, e.~g., the superconducting transition temperature, that will be discussed elsewhere.

\acknowledgements
We are grateful to A. V. Andreev, M. V. Feigel'man, L. I. Glazman, V. E. Kravtsov, P. M. Ostrovsky, I. V. Poboiko and K. S. Tikhonov  for stimulating discussions.



\appendix

\section{Friedel oscillations in the 2D case}
\label{app:2D}

In this Appendix, we illustrate how the ballistic DOS correction for a 2D metal
with a point-like interaction 
can be calculated using the approach developed in the main part of the paper. The purpose of this exercise is to reproduce the result of Rudin, Aleiner, and Glazman obtained in a slightly different technique \cite{RudinAleinerGlazman_TunnelingZBQuasiballistic_1997} and trace the difference between 2D and 3D Friedel oscillations responsible for a different behavior of $\delta\nu(E)$.

The real-space 2D Green function and its asymptotic behavior at $p_E r \gg 1$ have the form:
\be
\label{Green_function_2D}
	G_E^{R}(r) 
= - \frac{i m}{2} H^{(1)}_0 (p_E r)
\sim - \frac{m e^{i \pi / 4} } {\sqrt{2 \pi p_E r}} e^{i p_E r},
\ee
where $H^{(1)}_0$ is the Hankel function of the first kind. 
Since, contrary to the 3D case, the leading logarithmic contribution in 2D comes from $r \gg \lambda_F$ \cite{RudinAleinerGlazman_TunnelingZBQuasiballistic_1997}, we can replace the Green function by its asymptotics \eqref{Green_function_2D}.
Substituting it into Eqs.\ \eqref{analytical_continuation_F} and \eqref{analytical_continuation_H} and retaining the leading term in $p_Er\gg1$, we get
\be
\label{delta_nu_2D_sine_of_diff}
	\zeta_F - 2 \zeta_H = 
	\frac{4}{\pi^2}
	\int_{0}^{p_F}  \int_{0}^{\infty} dp_{E'}  dr 
	\, \sigma_{E'E}(r),
\ee
where $\sigma_{EE'}(r)$ is defined in Eq.\ \eqref{sigma-def} [a different order of energy arguments is due to a phase shift $\pi/4$ in the Green function \eqref{Green_function_2D}].
Regularizing the integral over $r$ in the infrared, 
one obtains
\be
\label{delta_nu_2D_singular}
	\zeta_F - 2 \zeta_H = -
	\frac{2}{\pi^2}
	\int_{0}^{p_F} dp_{E'} 
	\left( \frac{1}{p_E - p_{E'}} + \frac{1}{p_E + p_{E'}}  \right),
\ee
where the principal value of the integral is implied.

In the vicinity of the Fermi energy, at $|E - E_F| \ll E_F$, integration of the first term in Eq.~\eqref{delta_nu_2D_singular} yields the leading logarithmic contribution \cite{RudinAleinerGlazman_TunnelingZBQuasiballistic_1997}:
\be
\label{2D_logarithm}
	\frac{\delta \nu (E)} {\nu_0} = - \frac{1}{2 \pi}  \frac{\lambda}{k_F l}  \log \frac{|E - E_F|}{E_F} .
\ee
The second term in Eq.~\eqref{delta_nu_2D_singular} does not contribute to the leading part \eqref{2D_logarithm} since it does not contain a big logarithm at $E \rightarrow E_F$. Therefore no cancellation between the two terms is possible, contrary to the 3D case, where it is responsible for the vanishing of the correction at $E>E_F$, as discussed in Sec.~\ref{S:point-like}. Note also that the result \eqref{2D_logarithm} is particle-hole symmetric.

The above analysis relies on the assumption $p_Er\gg1$. 
In order to identify a relevant spatial scale responsible for the DOS correction, it is instructive to integrate in Eqs.\ \eqref{analytical_continuation_F} and \eqref{analytical_continuation_H} over $E'$ first, that leads to 
\be
\label{2D_integral_over_Eprime_first}
	\zeta_F - 2 \zeta_H = -
	\frac{4}{\pi^2}
	\int_{0}^{\infty} dr \,
	\frac{\sin 2 p_E r \sin 2 p_F r}{r} .
\ee
In the vicinity of the Fermi surface, $|E-E_F|\ll E_F$, this is a logarithmic integral coming from a broad range of lengths $1 \ll k_F r \ll E_F/|E-E_F|$. That justifies the approximation \eqref{Green_function_2D} as well as the stationary phase analysis of the scattering on Friedel oscillations performed in Ref.~\cite{RudinAleinerGlazman_TunnelingZBQuasiballistic_1997}.

\bibliography{dos_corrections}

\end{document}